\documentclass{desyproc}

\begin{document}
\title{Soft Scattering Re - Visited}
\def\pom{{I\!\!P}}
\def\beq{\begin{equation}}
\def\eeq{\end{equation}}
\def\bea{\begin{eqnarray}}
\def\eea{\end{eqnarray}}
\def\eq#1{{Eq.~(\ref{#1})}}
\def\fig#1{{Fig.~\ref{#1}}}
\newcommand{\bas}{\bar{\alpha}_S}
\newcommand{\as}{\alpha_S}
\newcommand{\bra}[1]{\langle #1 |}
\newcommand{\ket}[1]{|#1\rangle}
\newcommand{\bracket}[2]{\langle #1|#2\rangle}
\newcommand{\intp}[1]{\int \frac{d^4 #1}{(2\pi)^4}}
\newcommand{\mn}{{\mu\nu}}
\newcommand{\tr}{{\rm tr}}
\newcommand{\SP}{\langle \mid S \mid^2 \rangle}
\newcommand{\Tr}{{\rm Tr}}
\newcommand{\T} {\mbox{T}}
\newcommand{\braket}[2]{\langle #1|#2\rangle}
\newcommand{\Lb}{\left(}
\newcommand{\Rb}{\right)}
\setcounter{secnumdepth}{7}
\setcounter{tocdepth}{7}
\parskip=\itemsep
\newcommand{\lash}[1]{\not\! #1 \,}
\newcommand{\nn}{\nonumber}
\newcommand{\D}{\partial}
\newcommand{\h}{\frac{1}{2}}
\newcommand{\g}{{\rm g}}
\newcommand{\x}{\vec{x}}
\newcommand{\y}{\vec{y}}
\newcommand{\z}{\vec{z}}
\newcommand{\kv}{\vec{k}}
\newcommand{\qv}{\vec{q}}
\newcommand{\lv}{\vec{l}}
\newcommand{\mv}{\vec{m}}
\newcommand{\el}{{\cal L}}
\newcommand{\A}{{\cal A}}
\newcommand{\Ka}{{\cal K}}
\newcommand{\al}{\alpha}
\newcommand{\be}{\beta}
\newcommand{\ep}{\varepsilon}
\newcommand{\ga}{\gamma}
\newcommand{\de}{\delta}
\newcommand{\De}{\Delta}
\newcommand{\et}{\eta}
\newcommand{\ka}{\vec{\kappa}}
\newcommand{\la}{\lambda}
\newcommand{\ph}{\varphi}
\newcommand{\si}{\sigma}
\newcommand{\ro}{\varrho}
\newcommand{\tG}{\tilde{G}}

\author{{\slshape Uri Maor}\thanks{e-mail: maor@post.tau.ac.il}
\\[1ex]
Department of Particles Physics, School of Physics and Astronomy\\
Raymond and Beverly Sackler Faculty of Exact Science\\
Tel Aviv University, Tel Aviv 69978, Israel}

\contribID{maor_uri}

\maketitle

\begin{abstract}
An updated formulation of soft diffraction, compatible with $s$ and $t$
channel unitarity, is presented. Its consequent general soft scattering
features at high energies are explored.
The critical interplay between theory and data analysis and its 
consequent implications with regards to 
the theoretical foundations of soft scattering theory are discussed.
\end{abstract}
\section{Introduction}
\par
The renewed interest in soft scattering and Pomeron physics, which was dormant
for many years, can be traced to the market demand for reliable estimates of
hard diffraction gap survival probabilities, notably diffractive Higgs
production at the LHC. We note that the present vigorous study of
this subject is essentially based on sophisticated
utilization of relatively old theoretical ideas and
models, such as Gribov's Reggeon field theory\cite{Gribov},
Good and Walker (GW) decomposition of the proton wave function
accounting for low mass diffraction\cite{GW}
and the eikonal approximation\cite{eikonal} which secures the
compatibility of the scattering amplitudes with $s$ channel unitarity.
Compliance with $t$ channel unitarity is associated  with
multi Pomeron interactions ($\pom$ enhancement) which are a
generalization of
Mueller's triple Pomeron mechanism\cite{Mueller}, provided $G_{3\pom}$,
the triple Pomeron coupling, is not too small.
This mechanism supplements
GW diffraction with an additional high mass diffraction.
\par
In this talk I shall discuss the modeling and consequent soft scattering
predictions derived from the above dynamical considerations.
The implied gap survival probabilities will be discussed by Gotsman in the
following talk.
I shall assume a Regge like parametrization
in which the $\pom$ is super critical, i.e.
$\alpha_{\pom}(t)=1+\Delta_{\pom}+\alpha_{\pom}^{\prime}t$,
where $\Delta_{\pom}>1$.
The above $\pom$ exchange violates $s$-unitarity at high energies.
Recall that implementing $s$-unitarity is model dependent.
I shall confine myself to eikonal models which have the virtue of simplicity.
\par
In the ISR-Tevatron range $\sigma_{tot}$ and $\sigma_{el}$ are well
reproduced by Donnachie-Landshoff (DL) non screened Regge parametrization 
with $\Delta_{\pom}=0.08$ and $\alpha_{\pom}^{\prime}=0.25GeV^{-2}$.
The energy dependence of the soft diffractive cross sections
(notably $\sigma_{sd}$) is much milder, implying that strong screenings
initiated by $s$-unitarity must be taken into account.
As we shall see, the interplay between theory and data analysis results in
strong constraints on both $\Delta_{\pom}$ and
$\alpha_{\pom}^{\prime}$ inputs. This results in
profound consequences for the nature of the Pomeron and its QCD
foundations, providing a unifying interpretation of
soft and hard Pomerons.
\section{Good-Walker Eikonal Models}
\par
Consider a system of two Fock states, a hadron $\Psi_h$ and a diffractive
state $\Psi_D$ which are orthonormal.
The GW mechanism stems from the observation that these states do not
diagonalize the 2x2 interaction matrix ${\bf T}$.
Introduce two wave functions $\Psi_1$ and $\Psi_2$ which diagonalize
${\bf T}$,
\beq \label{2CHM}
A_{i,k}^{i',k'}=<\Psi_{i}\,\Psi_{k}|\mathbf{T}|\Psi_{i'}\,\Psi_{k'}>=
A_{i,k}\,\delta_{i,i'}\,\delta_{k,k'}.
\eeq
i.e. the $A_{i,k}$ amplitudes are constructed from the elastic
scatterings of $\Psi_{i}$ and $\Psi_{k}$.
In this representation the observed hadronic states are written
\beq \label{2CHM3}
\Psi_h=\alpha\,\Psi_1+\beta\,\Psi_2\,\,\,\,\,\,\,
\Psi_D=-\beta\,\Psi_1+\alpha \,\Psi_2\,,
\eeq
where $\alpha^2+\beta^2=1$.
The corresponding unitarity equations are
\beq \label{UNIT}
Im\,A^S_{i,k}\left(s,b\right)=|A^S_{i,k}\left(s,b\right)|^2
+G^{in}_{i,k}(s,b),
\eeq
where $G^{in}_{i,k}$ is the summed probability for
all non GW inelastic processes induced by an initial $(i,k)$ state.
A general solution of Eq.(\ref{UNIT}) can be written as
\beq \label{2CHM1}
A^S_{i,k}(s,b)=i \Lb 1 -\exp\Lb - \frac{\Omega^S_{i,k}(s,b)}{2}\Rb\Rb,
\eeq
\beq \label{2CHM2}
G^{in}_{i,k}(s,b)=1-\exp \Lb - \Omega^S_{i,k}(s,b) \Rb,
\eeq
where $\Omega^S_{i,k}$ are arbitrary.
In the eikonal approximation $\Omega^S_{i,k}$ are assumed to be real and
determined by the Born (non screened) input.
From Eq.(\ref{2CHM2}) we deduce that the probability that the initial projectiles
$(i,k)$ reach the final LRG diffractive interaction unchanged,
regardless of the initial state re-scatterings, is given by
$P^S_{i,k}=\exp {( - \Omega^S_{i,k}(s,b))}$.
In general, we have to consider four possible $(i,k)$ elastic
re-scattering options.
For initial $p$-$p$ (or $\bar p$-$p$) the two off diagonal amplitudes are
equal, $A^S_{1,2}=A^S_{2,1}$.
The corresponding elastic, SD and DD amplitudes are
\beq \label{EL}
a_{el}(s,b)=
i\{\alpha^4A^S_{1,1}+2\alpha^2\beta^2A^S_{1,2}+\beta^4\A^S_{2,2}\},
\eeq
\beq \label{SD}
a_{sd}(s,b)=
i\alpha\beta\{-\alpha^2A^S_{1,1}+(\alpha^2-\beta^2)A^S_{1,2}+\beta^2A^S_{2,2}\},
\eeq
\beq \label{DD}
a_{dd}=
i\alpha^2\beta^2\{A^S_{1,1}-2A^S_{1,2}+A^S_{2,2}\}.
\eeq
\par
The GW mechanism was originally conceived so as to describe a system
of a nucleon plus its diffractive $N^*$ isobars.
Obviously, this simplistic approach is not suitable for high
energy diffraction where $M^2_{diff}$
is bounded by 0.05$s$\footnote{this is an arbitrary bound commonly used.},
implying a continua of diffractive Fock states.
Throughout this talk I shall relate to GLMM\cite{GLMM} and
KMR\cite{KMR} models which are conceptually very similar, but differ
significantly in both their formalism and data analysis.
In the present context, two procedures were devised to overcome
the above difficulty: GLMM lump together all GW diffractive states
to an effective $|D>$ state,
to which we add the non GW $\pom$ enhanced high mass diffraction.
In this approach the GW contribution is very significant
and the mass distribution is smooth.
KMR and LKMR\cite{LKMR} chose to confine GW diffraction to
low $M_{diff}^2<10GeV^2$,
to which they add the high mass $\pom$ enhanced contribution.
In this approximation the bulk of the diffractive mass is non GW,
and its smoothness at $10GeV^2$ is not secured.
\par
GLMM, KMR and LKMR are multi channel eikonal models in which 
the initial re-scatterings of the incoming projectiles includes 
also diffractive states.
\beq \label{omega}
\Omega^S_{i,k}\Lb s,b \Rb = \nu^S_{i,k}\Lb s \Rb \Gamma^S_{i,k}\Lb s,b,... \Rb,
\eeq
where $\nu^S_{i,k}(s)=g_{i}g_{k}(\frac{s}{s_0})^{\Delta_{\pom}}$
and $\Gamma^S_{i,k}$ are the $b$-space profiles.
The profiles are external information in as much as, beside their
normalization and asymptotic constraints on their behavior,
they are determined by the data analysis.
In GLMM $\Gamma^S_{i,k}$ are given as
the $b$-transform of a two $t$-poles expression ($t=-q^2$).
Setting $\alpha_{\pom}^{\prime}$=0, the profiles are energy independent
\bea \label{prof}
&&\frac{1}{(1\,\,+\,\,q^2/m^2_i)^2}\times \frac{1}{(1\,\,+\,\,q^2/m^2_k)^2}
\,\,\,\Longrightarrow\,\,\,
\Gamma^S \Lb b; m_i, m_k;\alpha_\pom^{\prime} = 0 \Rb.
\label{SDB}
\eea
GLMM introduce a small energy dependence
\beq \label{MS}
m^2_i \,\,\,\Longrightarrow\,\,\,m^2_i(s)\,\,\equiv\,\,\,
\frac{m^2_i}{1\,\,+\,\,4m_{i}^2\alpha'_\pom\ln(s/s_0)}.
\eeq
The above parametrization is compatible with the requirements of
analyticity/crossing symmetry at large $b$, pQCD at large $q^2$
and Regge at small $q^2$. For details see Ref.\cite{GLMM}.
KMR and LKMR use a different parametrization for $\Gamma^S_{i,k}$
which is numerically compatible with GLMM.
The 3 groups reproduce ${d\sigma_{el}}/{dt}$ well in the
forward $t<0.5GeV^2$ cone.
\par
Consider a model in which diffraction is exclusively GW. This was recently
explicitly considered by GLMM and LKMR and non explicitly by KMR.
These, as well as earlier GW models. 
fit the (different) elastic sectors of their data bases, 
obtaining output fitted $\Delta_{\pom}=0.10-0.12$ with
${\chi^2}/{d.o.f.}<1.0$.
The above GW models fail to reproduce the diffractive sectors of their data
base. This deficiency is traced to the need to add the enhanced $\pom$
high mass contributions. This has been done in GLMM and KMR. LKMR
enhanced $\pom$ input is confined to Eq.(\ref{3P}).

\section{Multi Pomeron Interactions}
\par
Consider a single diffraction channel $p+p\rightarrow p+M_{sd}$.
Mueller's triple Pomeron mechanism, derived from 3 body unitarity,
leads to high SD mass which is non GW. In the leading order
\beq \label{3P}
M^2_{sd}\frac{d\sigma^{3\pom}}{dtdM^2_{sd}}\,=\,
\frac{1}{16\pi^2}g^2_p(t)g_p(0)G_{3\pom}(t)
\Lb\frac{s}{M^2_{sd}}\Rb^{2\Delta_{\pom}+2\alpha^{\prime}_{\pom}t}
\Lb\frac{M^2_{sd}}{s_0}\Rb^{\Delta_{\pom}}.
\eeq
The virtue of Eq.(\ref{3P}) is that $\Delta_{\pom}$ can be determined
from either the energy or mass dependences of the SD cross sections.
This approximation is valid for $s>>{M^2_{sd}}>>{m_p}$. 
\par
CDF analysis suggests\cite{CDF} a relatively large value of $G_{3\pom}$.
Consequently, we need to consider a very large family of multi
Pomeron interactions (enhanced $\pom$) which are not 
included in the GW mechanism.
As we shall see, this "new" dynamical feature initiates profound
differences in the calculated values of soft cross sections and
non GW diffractive gap survival probabilities (soft and hard).
Note that this features become significant above the Tevatron energy.
\par
GLMM and KMR treatment of $\pom$ enhanced interactions stems from Gribov
classical Reggeon calculus \cite{Gribov} and Kaidalov et al.\cite{Kaidalov}.
Recall that in this context the soft Pomeron is a simple pole in the
J-plane, while the hard (BFKL) Pomeron is a branch cut.
KMR model, which is a partonic model, derives directly from these sources.
Its summation is confined to semi enhanced
$\pom$ diagrams (see Fig.1b).  
KMR calculations are based on two ad hoc assumptions:
\newline
1) The coupling of a multi $\pom$ point vertex $n\pom \rightarrow m\pom$
($n+m>2$) is $g_m^n=\frac{1}{2}g_N nm\lambda^{n+m-2}$. In this notation
$G_{3\pom}=\lambda g_N$. Note that in Kaidalov et al.
$g_m^n=\frac{1}{2}g_N \lambda^{n+m-2}$.
\newline
2) Most of LHC non GW diffractive reactions of interest are hard.
Given a $3\pom$ vertex, $G_{3\pom}$ is unchanged
by the interchange of soft and hard Pomerons. This is not self evident. 
A possible support for the above is obtained from GLMM interpretation 
of the Pomeron (see below).
\par
\begin{wrapfigure}{r}{0.45\textwidth}
\includegraphics[width=65mm,height=30mm]{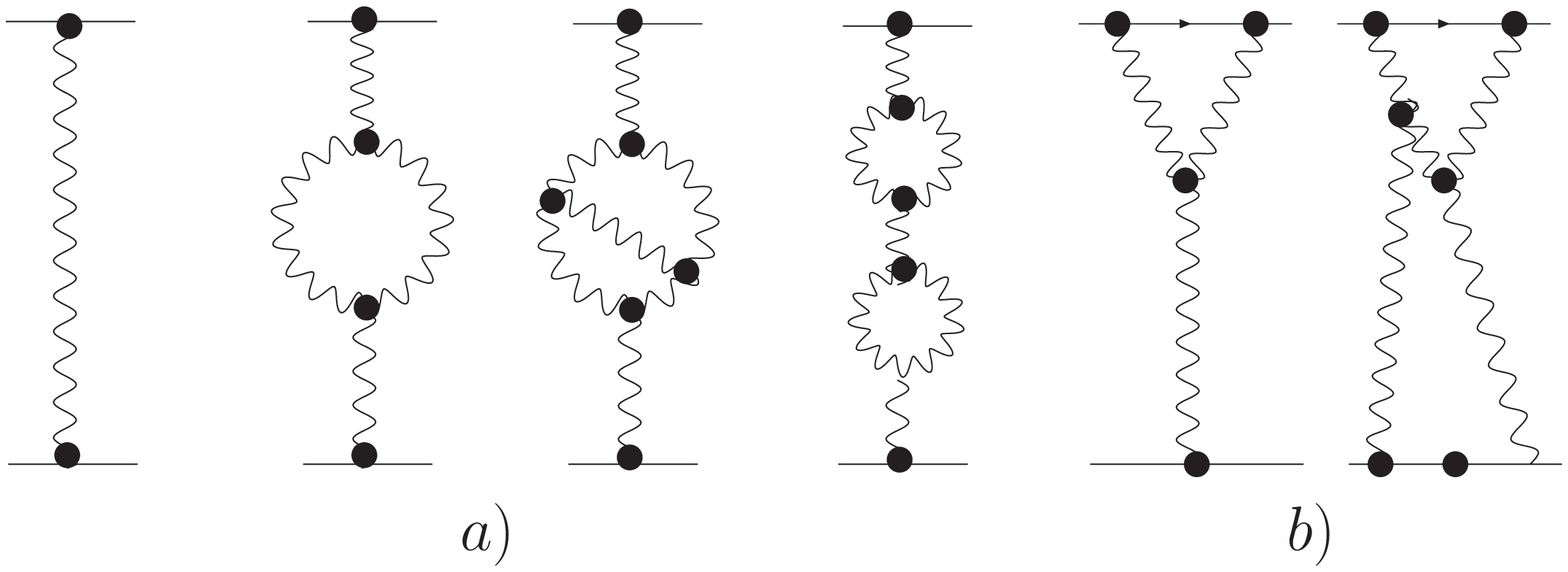}
\caption{Low order terms of the Pomeron Green's
function. a) Enhanced. b) Semi-enhanced.}\label{enh}
\end{wrapfigure}
As we shall see in the next chapter, the data analysis executed by GLMM and KMR 
converges to compatible exceedingly small $\alpha^{\prime}_{\pom}$ values  
and high, BFKL like, $\Delta_{\pom} \simeq 0.30-0.35$. 
The adjustments of these parameters are correlated. In non screened Regge model
$\Delta_{\pom}$ controls the elastic cross section energy dependence, while 
$\alpha^{\prime}_{\pom}$ controls the energy dependence of the 
(shrinking) elastic slope. 
As $\alpha^{\prime}_{\pom}$ gets smaller $\Delta_{\pom}$ 
becomes larger initiating stronger screening which compensates the reduction 
of $\alpha^{\prime}_{\pom}$, and vice versa. As we saw the vanishing value of the 
fitted $\alpha^{\prime}_{\pom}$ was explicitly dictated by the b-profiles GLMM 
and KMR chose. Non explicitly this choice initiated also the high 
$\Delta_{\pom}$ fitted value. These results 
have profound implications:
\newline  
1) A key observation of GLMM is that the exceedingly small fitted value
of $\alpha_{\pom}^{\prime}$ implies that the "soft" $\pom$ is hard enough to be
treated perturbatively. Following Gribov we identify the correlation between
$\alpha_{\pom}^{\prime}$ and $<p_t>$, the mean transverse momentum of the
partons (actually, color dipoles) associated with the $\pom$.
$<p_t>=1/\sqrt{\alpha_{\pom}^{\prime}}$, from which we deduce that the
QCD running coupling constant $\alpha_S<<1$. Accordingly, we proceed from 
Gribov's parton model interpretation to pQCD. 
GLMM sum over the enhanced diagrams. 
Technically, we have adopted the MPSI procedure\cite{MPSI} 
in which $g_m^n$ is reduced to a sequence of triple $\pom$
vertexes (Fan diagrams). For details see Ref.\cite{GLMM}.
This may pose a problem
for the calculation of SD cross section for which the lowest order
diagram is semi enhanced. To avoid this problem we have added to this 
calculation a term by term 
summation of the relevant semi enhanced diagrams.
\newline
2) The fitted high value of $\Delta_{\pom}$ initiates strong screening which 
results in a renormalization of the Pomeron exchange amplitudes. As a result 
$\Delta^{eff}_{\pom}$ is reduced monotonically with energy. 
In GLMM calculations 
$\Delta^{eff}_{\pom}(2TeV) \simeq 0.070$, 
$\Delta^{eff}_{\pom}(14TeV) \simeq 0.045$ and 
$\Delta^{eff}_{\pom}(60TeV) \simeq 0.032$. 
KMR results are compatible with ours, see Table 2. 
The slow decrease of $\Delta^{eff}_{\pom}$ raises the question if 
its value may become negative at high enough energies, larger than 
$W=10^5 GeV$ which is the bound of validity of both GLMM and KMR. 
In GLMM we have checked that $\Delta^{eff}_{\pom}(100TeV) > 0$.  
I am less clear about KMR. 
The compatibility between GLMM and KMR is surprising. As noted, GLMM 
sum over the enhanced diagrams (Fig.1a) while KMR sum over the semi enhanced 
diagrams (Fig.1b). Very intuitively (at the risk of being wrong), it seems that 
GLMM renorrmalize the $\pom$ propagator while KMR renorrmalize the $\pom$ vertex. 
A complete calculation should, obviously, include both contributions. 
\section{The Interplay Between Theory and Data Analysis}
The goal of the data analysis of interest   
is to adjust a set of theoretical parameters.
To this end we construct: 
\newline 
1) A suitable data base adjusted, with a satisfactory resolution, 
to fix the parameters.  
\newline
2) An adjustment procedure, be it a fit (GLMM, LKMR) or 
tuning by trial and error (KMR). 
We distinguish between 
a parameter adjustment which is executed through a 
reconstruction of the entire data base in one step (GLMM),
and a sequence of factorizable adjustment steps (KMR, LKMR), 
each adjusting a sub group of the free parameters.
\newline
There is a significant difference between the data analysis carried out by 
GLMM and KMR. This reflects both in the choices of data bases made by the
two groups and their adjustment procedures.
The starting point of both investigations is the
observation that a GW model reproduces the elastic data well, but 
its reproduction of the diffractive sector is deficient.
Both groups claim to achieve an improved
reproduction of their over all data base once the contributions
of enhanced Pomeron diagrams are included. 
\par
GLMM have constructed a global data base
so as to fit the multitude of free parameters. 
It includes $\sigma_{tot}$, $\sigma_{el}$, $\sigma_{sd}$, 
$\sigma_{dd}$ and $B_{el}$ in the ISR-Tevatron range, 
CDF differential elastic cross sections and SD mass distribution 
were checked for consistency.
The conceptual approach of KMR and LKMR is completely different.
Their data base contains only the measured values of
$d\sigma_{el}/dt$, which enables to predict $\sigma_{tot}$,
and $d\sigma_{sd}/dtd(M^2_{sd}/s)$.
In my opinion KMR data base is too limited to enable a substantiation
of their goals. Specifically:
\newline
1) As we saw, the $b$-profiles $\Gamma^S_{i,k}$ control the
features of $d\sigma_{el}/dt$ which are only weakly
coupled to the proposed dynamics.
Indeed, 6 models, published over the last 10 years (3 KMR+LKMR+2 GLMM) 
with different dynamics, 
i.e., exclusive GW (GLMM, LKMR), GW+zero order $\pom$ enhancement (LKMR) 
and GW+$\pom$ enhancement (GLMM, KMR). The output parameters spread over   
$0.1 \leq  \Delta_{\pom} \leq 0.55$ 
and $0 \leq \alpha_{\pom}^{\prime} \leq 0.066$. All 6 models 
reproduce, almost identically, the CDF
distributions of $d\sigma_{el}/dt$. 
The unavoidable conclusion is that a reconstruction of $d\sigma_{el}/dt$ 
on it own has no resolution power. 
The only common ingredient to all 6 models is their compatible b-profiles. 
These profiles constrain $\alpha_{\pom}^{\prime}$ to very small values. 
This is the key observation leading to a   
pQCD (GLMM) or partonic (KMR) $\pom$ interpretation.
\newline 
2) The reconstruction of CDF 
$d\sigma_{sd}/dtd(M^2_{sd}/s)$ by GLMM, LKMR and KMR 
partially substantiates the introduction of high mass 
multi Pomeron interactions. In my opinion, this investigation, 
in its present state state, is unable to provide a decisive 
verdict on this issue. GLMM and KMR sum different sectors of the 
enhanced diagrams. LKMR take into account only the lowest order 
Mueller's $3\pom$ diagram. Regardless of these differences, the 
three groups produce compatible output results which indicate 
that CDF data, as is, is not sufficient to differentiate between 
different modes of $\pom$ enhanced diagram summations.
\newline
3) To further clarify the experimental limitations, 
let us recall that CDF define their high mass diffraction bound 
$1 - x_L = \frac {M^2_{sd}}{s} \leq 0.15$ (the common bound is 0.05).
KMR and LKMR 
define a lower bound  $M^2_{sd} > 10 GeV^2$ which corresponds to 
$1 - x_L = 3 \cdot 10^{-6}$. 
Note that CDF SD mass distribution available for analysis corresponds to 
$1 - x_L > 2 \cdot 10^{-3}$. This mass distribution covers less than 20$\%$ of 
KMR expected high diffractive mass. GLMM have a completely different 
classification in which the GW low mass diffraction reduces monotonically 
with no arbitrary upper bound and the high mass is defined above 
$10 GeV^2$, identical to KMR. Additional difficulties with the analysis 
of CDF data with $1 - x_L > 0.02$ is that an arbitrary significant 
background contribution has to be added. It is induced by secondary Regge 
diagrams such as $\pom \pom R$. An added element of ambiguity is  
that LKMR with a zero order $3\pom$ calculation is faring as 
well as the high order summations of GLMM and KMR.
\par
\begin{table}[h]
\begin{tabular}{|l|l|l|l|l|l|l|l|l|}
\hline
& & &$\alpha_{\pom}^{\prime}$&
$g_1$ & $g_2$ &
$m_1$ & $m_2$ & \\
& $\Delta_\pom$ & $\beta$ &
$GeV^{-2}$ & $GeV^{-1}$ & $GeV^{-1}$ &
$GeV$ & $GeV$ & $\chi^2/d.o.f.$ \\
\hline
\,\,\,\,\,\,\,GW\,\,&0.120 & 0.46 & 0.012  &
1.27  & 3.33  & 0.913 & 0.98  &
\,\,\,\,\,\,0.87 \\
\hline
\,GW+$\pom-enh.$  & 0.335 & 0.34 &  0.010 & 5.82  &
239.6  & 1.54 & 3.06  &
\,\,\,\,\,\,1.00 \\
\hline
\end{tabular}
\caption{Fitted parameters for GLMM GW and GW+$\pom$-enhanced models.}
\label{t1}
\end{table}
\par 
GLMM fitting procedure aims to reproduce our global data base. 
A fit with a GW model (no $\pom$-enh) provides excellent reproduction of our
elastic sector while the the reproduction of the diffractive sector is very poor!
The repeated fit with a GW+$\pom$-enh model results with a very good $\chi^2$.
The outputs of both models are presented in Table 1. Checking we note 
that the exceedingly small value of $\alpha^{\prime}_{\pom}$ is persistently 
obtained in both models. The outputs of $\Delta_{\pom}$ and $g_2$ change 
drastically once $\pom$-enh is included. 
As we shall see this has significant consequences for the 
approach of $a_{el}(s,b)$ toward the black disc bound.   
As we have noted, KMR and LKMR tune $g_1$ and $g_2$ through a reproduction of 
$d\sigma_{el}/dt$ and the other parameters through a reproduction of 
$d\sigma_{sd}/dtd(M^2_{sd}/s)$. 
The values they obtain for $\Delta_{\pom}$ and 
$\alpha^{\prime}_{\pom}$ are compatible with ours. 
\begin{table}[ht]
\begin{tabular}{|l|l|l|l|}
\hline
& \,\,\,\,\,\,\,\,\,\,\,\,\,\,\,\,\,\,\,Tevatron
& \,\,\,\,\,\,\,\,\,\,\,\,\,\,\,\,\,\,\,\,\,\, LHC
& \,\,\,\,\,\,\,\,\,\,\,\, W=$10^5$ GeV  \\
& GLMM\,\,KMR(07)\,\,KMR(08)
& GLMM\,\,KMR(07)\,\,KMR(08)
& GLMM\,\,KMR(07)\,\,KMR(08) \\
\hline
$\sigma_{tot}$ 
&\,\, 73.3 \,\,\,\,\,\,\,\,\,\,74.0\,\,\,\,\,\,\,\,\,\,\,\,\,\,73.7
& \,\,\,\,92.1
\,\,\,\,\,\,\,\,\,\,\,88.0\,\,\,\,\,\,\,\,\,\,\,\,\,\,\,91.7
&\,108.0 \,\,\,\,\,\,\,\,\,\,\,98.0\,\,\,\,\,\,\,\,\,\,\,\,\,108.0 \\
\hline
$\sigma_{el}$ 
& \,\, 16.3 \,\,\,\,\,\,\,\,\,\,16.3\,\,\,\,\,\,\,\,\,\,\,\,\,\,16.4
& \,\,\,\,20.9
\,\,\,\,\,\,\,\,\,\,\,20.1\,\,\,\,\,\,\,\,\,\,\,\,\,\,\,21.5
& \,\,\,\,24.0 \,\,\,\,\,\,\,\,\,
22.9\,\,\,\,\,\,\,\,\,\,\,\,\,\,\,\,26.2 \\
\hline
$\sigma_{sd}$ 
& \,\,\,\,\, 9.8 \,\,\,\,\,\,\,\,\,\,10.9\,\,\,\,\,\,\,\,\,\,\,\,\,\,13.8
& \,\,\,\,11.8
\,\,\,\,\,\,\,\,\,\,\,13.3\,\,\,\,\,\,\,\,\,\,\,\,\,\,\,19.0
& \,\,\,\,14.4
\,\,\,\,\,\,\,\,\,\,\,15.7\,\,\,\,\,\,\,\,\,\,\,\,\,\,\,\,24.2 \\
\hline
$\sigma_{dd}$ & \,\,\,\,\, 5.4\,\,\,\,\,\,\,\,\,\,\,\,\,\,\,7.2
& \,\,\,\,\,\,\,6.1 \,\,\,\,\,\,\,\,\,\,\,13.4
& \,\,\,\,\,\,\,6.3\,\,\,\,\,\,\,\,\,\,\,\,\,17.3 \\
\hline
\end{tabular}
\caption{Comparison of GLMM, KMR(07) and KMR(08) cross sections in $mb$.}
\label{T}
\end{table}
\par
GLMM and KMR high energy Tevatron, LHC and Cosmic Rays predicted
cross sections are summarized in Table 2.
The elastic and total cross section outputs of the two models are compatible
and, above the Tevatron, significantly lower than those
obtained in models with no multi-Pomeron contributions.
This is a consequence of $\Delta_{\pom}$ renormalization due to the enhanced
$\pom$ contributions.
GLMM and KMR(07) predicted $\sigma_{sd}$ are compatible,  
where KMR(07) are consistently larger by approximately 12$\%$.
KMR(08) predicted $\sigma_{sd}$ are considerably larger than GLMM 
and are growing at a faster rate.
The difference between KMR(07) and GLMM $\sigma_{dd}$ predictions is even 
more dramatic, where $\sigma_{dd}(KMR(07)/\sigma_{dd}(GLMM) \simeq 3$ 
at $W=10^5 GeV$. This very large difference is 
due to KMR large diffractive high mass predictions (see Table 2).
The recent KMR(08) neglects to mention
the high diffractive mass sector of double diffraction while showing higher
SD cross sections than in KMR(07).
\par
In my opinion the GLMM and KMR compatible predictions
of total and elastic cross sections
at the LHC and AUGER are of fundamental importance because they are
significantly lower than the predicted values based on unitarity models with no
$\pom$ enhancement. These measurements may
provide a decisive support for the importance of multi-Pomeron interactions
at high enough energies
and, consequently, imply that the growth of the total and elastic cross
sections with energy is much more moderate than anticipated.
This feature reflects in the slow decrease of $\Delta_{\pom}^{eff}$ 
shown in Sec 3. 
\section{The Approach Toward the Black Disc Bound}
The GW base amplitudes of GLMM are $A^S_{i,k}$, 
with $b$ dependences specified in \eq{prof} - \eq{MS}.
These are the building blocks
with which we construct $a_{el}$, $a_{sd}$
and $a_{dd}$ (\eq{EL} - \eq{DD}).
The $A^S_{i,k}$ amplitudes are bounded by the black
disc unitarity limit of unity.
Checking GLMM fitted parameters, presented in Table 1,
we observe that $g_1$ and $g_2$, which are comparable in
the GW model, significantly change in the GW+$\pom$ enhanced model
where we obtain $g_2>>g_1$. 
The implication of of our fitted values of $g_1$ and $g_2$, is that including 
the diffractive data in our global fit forces a large 
inequality between the three GW $A^S_{i,k}$ components.
$A^S_{2,2}(s,b=0)$ reaches unity at a very low energy, 
$A^S_{1,2}(s,b=0)$ reaches unity at approximately W=100GeV 
and $A^S_{1,1}(s,b=0)$ reaches unity at exceedingly high energies, 
well above LHC.
The observation that one, or even two, of our $A^S_{i,k}(s,b)$=1 does not
imply that the elastic scattering amplitude 
has reached the unitarity bound at these $(s,b)$ values.
$a_{el}(s,b)$ reaches the black disc bound when, and only when,
$A^S_{1,1}(s,b)$=$A^S_{1,2}(s,b)$=$A^S_{2,2}(s,b)$=1, independent of $\beta$.
The approach of $a_{el}(s,b=0)$
toward the black bound depends on the rate of
$A^S_{1,1}(s,b)$ increase. This rate is very slow because 
of the relative smallness of $g_1$ to which we add that 
$A^S_{1,1}$ increase with energy above LHC  
becomes ever so moderate as a consequence of
the renormalization reduction of $\Delta_{\pom}$.
Our results are different from the predictions of most 
available models, notably KMR,  
in which $a_{el}(s,b=0)$ reaches unity a few $TeV$ above LHC. Note, though,   
that GLMM is the only model which includes the diffractive 
along side the elastic data in its data analysis. 
All models which predict saturation of $a_{el}(s,b=0)$ just above LHC 
have confined their data analysis exclusively to the elastic sector. 
\par
A consequence of the input $\Omega^S_{i,k}$ being very large at small $b$,
is that $P_{i,k}^S(s,b)$ is exceedingly small at these small $b$ values. 
As a result, given a diffractive (non screened) input,
its output (screened)
amplitude is peripheral in $b$. This is a general feature, common to
all eikonal models regardless of their b-profiles details.
The general behavior indicated above
becomes more extreme at ultra high energies,
where $a_{el}$ continues to get darker and expand.
Consequently, the inelastic diffractive channels (soft and hard) 
becomes more and more
peripheral and relatively smaller when compared with the elastic channel.
Given $(s,b)$ at which $a_{el}(s,b)$ = 1,
the corresponding diffractive amplitudes, GW and non GW, vanish. 
\par 
The behavior of the ratio 
$R_D=(\sigma_{el}+\sigma_{sd}+\sigma_{dd})/\sigma_{tot} $ 
conveys information regarding the onset of s-unitarity at very 
high energies. Assuming diffraction to be exclusively GW, 
we obtain\cite{Pumplin} $R_D \leq 0.5$. Multi $\pom$ induced diffracion is not 
included in $R_D$ since it originates from $G^{in}_{i,k}$. 
Hence their non screened high mass cross section is suppressed by 
its survival probability which decreases with energy. 
In GLMM $R_D<0.5$, decreasing slowly. In KMR(07) $R_D>0.5$, 
increasing slowly with energy. The partial information available 
on KMR(08) suggests that its $R_D$ grows even faster.  
\section{Conclusions}
This presentation centered on the phenomenology implied by 
multi Pomeron dynamics incorporated in soft diffraction and 
its consequences on soft scattering.
\par
The concept of $\pom$ enhancement was triggered, at the time, by the 
assessment that $G_{3\pom}$ is not too small. Our view of the Pomeron got more 
focused with the updated data analysis of soft scattering in which we get   
$\Delta_{\pom} \simeq 0.30-0.35$ and  $\alpha_{\pom}^{\prime} \simeq 0.01$. 
The implied KMR Pomeron is hard enough to be treated partonically, in which 
the traditional classification of the soft $\pom$ as a simple J-pole and the 
hard $\pom$ as a branch cut in the J-plane is maintained. GLMM went further on  
identifying the soft $\pom$ with the hard $\pom$. This is, clearly, a fundamental 
theoretical issue which should be further investigated.
\par
As it stands this dynamics is compatible with the data, but at present we 
can not support it with a decisive verification. 
A prediction shared by GLMM and KMR is the expected significant reduction, 
compared with non screened predictions, of $\sigma_{tot}$ 
and $\sigma{el}$ at the LHC. In my opinion this measurment is of a 
critical value.  
\par
Decisive experimental signatures of $\pom$-enh are expected, 
essentially, above the Tevatron.
Consequently, we should be prudent when evaluating phenomenological models 
which reproduce the Tevatron data well. This is, obviously, required of a 
successful model, but is definitely not sufficient.
\section{Acknowledgments}
This research was supported in part by BSF grant $\#$ 20004019
and by a grant from Israel Ministry of Science 
and the Foundation for Basic Research
of the Russian Federation.

\begin{footnotesize}

\end{footnotesize}

\end{document}